\documentclass{aa}
\usepackage{graphicx}

\def\gtsim{\raise 2pt \hbox {$>$} \kern-1.1em \lower 4pt \hbox {$\sim$}}
\def\ltsim{\raise 2pt \hbox {$<$} \kern-1.1em \lower 4pt \hbox {$\sim$}}
\def\ltapprox{\raise 2pt \hbox {$<$} \kern-1.1em \lower 5pt \hbox {$\approx
$}}
\def\gtsim{\raise 2pt \hbox {$>$} \kern-1.1em \lower 4pt \hbox {$\sim$}}
\def\gtapprox{\raise 2pt \hbox {$>$} \kern-1.1em \lower 5pt \hbox {$\approx
$}}

\begin{document}
%
%
%
\title{Spectral properties and origin of the radio halo in A3562}
%
%
\author
{
S.~Giacintucci\inst{1,}\inst{2,}\inst{3} \and
T.~Venturi\inst{3} \and
G.~Brunetti\inst{3} \and
S.~Bardelli\inst{1} \and
D.~Dallacasa\inst{2,}\inst{3} \and
S.~Ettori\inst{1} \and
A.~Finoguenov\inst{4,} \and
A.~P.~Rao\inst{5} \and
E.~Zucca\inst{1}
}
\institute
{
INAF -- Osservatorio Astronomico di Bologna, 
via Ranzani 1, I--40127 Bologna, Italy
\and
Dipartimento di Astronomia, Universit\`a di Bologna,
via Ranzani 1, I--40127, Bologna, Italy
\and
INAF -- Istituto di Radioastronomia, via Gobetti 101, I-40129, Bologna, Italy 
\and
Max--Planck--Institut f\"ur extraterrestrische Physik, 85740 Garching, Germany
\and
National Centre for Radio Astrophysics (TIFR), Pune University Campus, Post Bag No.3,
Ganeshkhind, Pune 411 007, India
}
%
\date{Received 00 - 00 - 0000; accepted 00 - 00 - 0000}
%
\titlerunning{The origin of the radio halo in A3562}
\authorrunning{Giacintucci et al.}
\abstract{We present a new detailed multiband study of the merging
cluster A3562, in the core of the Shapley Concentration Supercluster.
We analyzed new, low frequency radio data performed at 240 MHz, 
332 MHz and 610 MHz with the Giant Metrewave Radio Telescope (GMRT).
The new GMRT data allowed us to carry out a detailed study of
the radio halo at the centre of A3562, as well as of the head--tail
radio galaxy J1333--3141 embedded in it, and of the  
extended emission around the peripheral cluster galaxy J1332--3146a.
Thanks to the present observations we could derive the integrated spectrum 
of the radio halo with five data points in the frequency range 240 MHz --
1.4 GHz. Our data show a clear steepening of the total spectrum 
in this frequency range.
Furthermore, by comparing the GMRT 332 MHz image with a previously published
VLA 1.4 GHz image, we produced an image of the halo spectral index distribution.
The image shows a very complex structure, with an average value of
$\alpha^{1.4GHz}_{332MHz} \sim 1.5$ and a number of knots steepening up to 
$\sim 2$.
We performed a combined morphological and statistical analysis using the
radio images and the quantities 
derived from XMM--Newton and {\it Chandra} observations.
We discuss our results in the light of particle re--acceleration processes in
galaxy clusters. 
In particular, we outline an overall picture, consistent
with the available radio and X--ray data, in which the cluster merger 
kinematics, the injection of turbulence and B--amplification induced by 
the merger between A\,3562 and SC\,1329--313 are jointly taken into account.
}
\maketitle
%
%
\section{Introduction}

Cluster radio halos belong to a particular class of  
steep--spectrum ($\alpha>1$, S$\propto\nu^{-\alpha}$) radio sources, 
which are identified with regions of diffuse emission 
on Mpc scale, with no obvious optical counterpart.
They are located at the centre of X--ray luminous and 
massive galaxy clusters (for a recent review see Feretti \cite{feretti03}),
and their number is steadily growing.
They are produced by synchrotron radiation from a population of 
relativistic electrons, coexisting with the thermal electrons and protons
in the intracluster medium (ICM).  
Their origin seems to be strictly connected to the X--ray 
properties of the hosting cluster and to the presence of
cluster merger activity (Buote \cite{buote01};
Schuecker et al. \cite{schuecker01}), since radio halos have been found 
only in clusters with significant substructure both in the galaxy distribution and
X--ray brightness. 

Merging processes can generate turbulence, which may
amplify the cluster magnetic fields and re--accelerate 
the relativistic electrons diffused in the ICM, producing the emission 
we observe from radio halos (see e.g. the review papers by 
Brunetti 
\cite{brunetti03}, Brunetti \cite{brunetti04}, and Sarazin \cite{sarazin04}).
The connection between the relativistic plasma and the thermal ICM 
is supported by the data available thus far, 
which suggest the existence of a correlation between the halo radio 
power and the X--ray luminosity, temperature and mass 
of the hosting clusters (Colafrancesco et al. \cite{colafrancesco99};
Liang et al. \cite{liang00}; 
Govoni et al. \cite{govoni01a}; Bacchi et al. \cite{bacchi03}).
In particular, in a number of clusters with radio halo, 
Govoni et al. (\cite{govoni01b}) showed that the extent and 
the shape of the halo emission closely follow the X--ray structure 
of the cluster on large scales. In particular higher radio brightness is 
associated with higher X--ray brightness of the gas.

The radio halo studied in this paper is located in the 
galaxy cluster A3562, in the core region of the Shapley
Concentration Supercluster, at a mean redshift of $z$= 0.05.
A3562 belongs to the A3558 complex, one of the most  
spectacular examples of cluster merger. The merging stage of this chain of 
three clusters (A3558, A3562 and A3556) and two groups of galaxies 
(SC\,1327--312 and SC\,1329--313) has been extensively outlined by studies in the optical  
and X--ray bands (Bardelli et al. \cite{bardelli96}, \cite{bardelli98a}, \cite{bardelli98b},
 \cite{bardelli00} and \cite{bardelli02}; Ettori et al. \cite{ettori97} and \cite{ettori00}), as
 well as in the radio band (Venturi et al. \cite{venturi97}, \cite{venturi98},
 \cite{venturi99}, \cite{venturi00} and \cite{venturi03}, hereinafter
 V00 and V03 respectively; Giacintucci et al. \cite{giacintucci04}, hereinafter G04).
 Recently, Finoguenov et al. (\cite{finoguenov04},
hereinafter F04) carried out a detailed study of the hydrodynamic state 
of A3562 and the nearby companion SC\,1329--313 through XMM--Newton observations.
They showed that the region of A3562 and SC\,1329--313 is clearly 
disturbed, suggesting a recent interaction between the two clusters. 
In particular F04 proposed that SC\,1329--313 has recently
($\sim 1$ Gyr ago) passed North of A3562, inducing a
sloshing of the cluster core in the North--South direction.

The A3562 halo is one of the smallest and the less powerful 
radio halos known to date, with a largest linear size of $\sim470$ kpc and a total radio power 
at 1.4 GHz of P$_{1.4GHz}$ $\sim$1.1 $\times$ 10$^{23}$ 
W Hz$^{-1}$ (V03).\footnote{Throughout the paper we assume $H_0 = 70$ km s$^{-1}$ Mpc$^{-1}$,
$\Omega_m$=0.3 and $\Omega_{\Lambda}$=0.7. At the average redshift of the
A3558 complex ($z=0.05$), this cosmology leads to a linear scale of 
1 arcsec=0.98 kpc.} 
The X--ray properties of A3562, such as the luminosity and temperature, are 
less extreme than those of the other clusters hosting a radio halo.
In particular, kT=5.1$\pm$0.2 keV, and L$_{X,bol} = 2.2\times10^{44}$ erg s$^{-1}$
(Ettori et al. 2000).
For this reason the halo source in A3562 plays a special role in understanding 
the origin of radio halos and their connection to the X--ray properties of 
the intergalactic medium and the merging activity of the hosting cluster. 
As shown in V03, the A3562 halo follows and extends the correlations 
between the halo radio power and the cluster X--ray luminosity and temperature 
down to lower values of all quantities involved. This suggests that A3562--like halos 
may help us to understand the missing link between the massive and X--ray luminous clusters
with a radio halo and those without (or undetected). 

In this paper we present new low frequency radio data
of A3562, obtained with the Giant Metrewave Radio 
Telescope (GMRT) at 240 MHz, 332 MHz and 610 MHz. 

The observations are centered on the core of A3562 and cover also the region of 
the nearby group SC\,1329--313, East of A3562, where a region of diffuse 
radio emission extends around the radio galaxy J1332--3146a, at the 
border of the X--ray emission of the group (G04). 

In Sect. 2 and 3 we present the new GMRT observations. 
In Sect. 4 we combine the new data with the 1.4 GHz VLA data 
from V03, to discuss the morphological and spectral properties of the radio 
halo. The total radio spectrum of the halo is modelled in Sect. 5. In Sect. 6
we perform a combined radio/X--ray analysis of the A3562 region,
using XMM--Newton (F04) and {\it Chandra} (Ettori et al. in preparation) 
observations. Finally, in Sect. 7 we propose a cluster merger scenario
to account for the observed properties of A3562.

\section{GMRT low frequency radio observations}
%
%
\begin{table*} 
\caption[]{Details of the GMRT observations for A3562 and SC\,1329--313.}
\begin{center}
\begin{tabular}{ccccccccc}
\hline\noalign{\smallskip}
Frequency & Bandwidth  & RA$_{J2000}$ & DEC$_{J2000}$ & Primary beam  & FWHM, PA   & {\bf u$-$v range} & Obs. time &  rms     \\
  MHz &  MHz   &  h,m,s  & $\circ$, $\prime$, $\prime \prime$ &   $\prime$      &  (full array)  &   {\bf k$\lambda$} & h       &    mJy b$^{-1}$ \\
          &            &            &                               &                   & $\prime \prime$, $\circ$ &        &   &                         \\
\\
\noalign{\smallskip}
\hline\noalign{\smallskip}
240   &   8     &  13 33 30 & -31 41 00 & 108    & 17.4 $\times$ 11.5, 11&    {\bf $\sim 0.05 - 20$} & 7   &  0.80     \\
332   &   16     &  13 33 30 & -31 41 00 &  87    & 15.0 $\times$ 8.0, 26 &  {\bf $\sim 0.10 - 30 $} &  7   & 0.30 \\
610   &   8     &  13 33 30 & -31 41 00 &  43    &  7.3 $\times$ 5.00, 7 &    {\bf $\sim 0.15 - 50$} & 7   &  0.15\\

\noalign{\smallskip}
\hline
\end{tabular}
\end{center}
\label{tab:obs}
\end{table*}
%
%

We observed the region of A3562 and SC\,1329--313 using the 
GMRT simultaneously at 240
and 610 MHz on 15 January 2003, and at 332 MHz on 19 February 2003,
for a total of $\sim$ 7 hours at each frequency.

The observations were carried out in spectral line mode with
64 channels at 240/610 MHz and 128 channels at 332 MHz,
with a spectral resolution of 125 kHz.

The details of the observations are summarized in Table \ref{tab:obs}.

The data calibration and reduction were performed using the
the NRAO Astronomical Image Processing System (AIPS) 
package. 
At 240/610 MHz the sources 3C147 and 3C286 were used to determine 
and correct for the bandpass shape and for the initial 
amplitude and phase calibration . 
At 332 MHz only 3C286 was used as primary calibrator. 
The secondary phase calibrators were 1248--199 and 1311--222
at 240/610 MHz, 1311--222 at 332 MHz.

An accurate editing was needed to
identify and remove the data affected by radio frequency 
interference (RFI), especially at 332 MHz and 240 MHz. 
In order to find a compromise between the size
of the data sets and the need to minimize bandwidth smearing 
within the primary beam, the central channels were averaged 
using the AIPS task SPLAT to 6 channels of 1 MHz each at 
240 MHz and 2 MHz each at 332 MHz. At 610 MHz the 
central band was averaged to 1 single channel of 6 MHz.

After some steps of phase self--calibration, at each frequency we produced images 
in a wide range of resolutions using the wide--field imaging 
technique, with 25 facets covering a field of view in the range 
$1.4^2 \div 3.5^2$ deg$^2$, depending on the frequency and resolution.
The final images were 
combined with the task FLATN and corrected for the primary beam 
appropriate to the GMRT antennas. As a final step, since right ascension and declination 
of the final fields were referred to the observation epoch, we 
converted them to the J2000 epoch by means of the AIPS task REGRD. 

The average noise achieved in the final full resolution and tapered images 
is $\sim 0.80 $ mJy b$^{-1}$ at 240 MHz, $\sim 0.30 $ mJy b$^{-1}$ at 332 MHz 
and $\sim 0.15 $ mJy b$^{-1}$ at 610 MHz. These values are a few times higher
than the expected thermal noise. 
The residual amplitude calibration errors are of the order of 10\% at 610 MHz,
20\% at 332 MHz, and 15\% at 240 MHz.

%
%
\begin{figure*}
\centering
\includegraphics[angle=0,width=5.9cm]{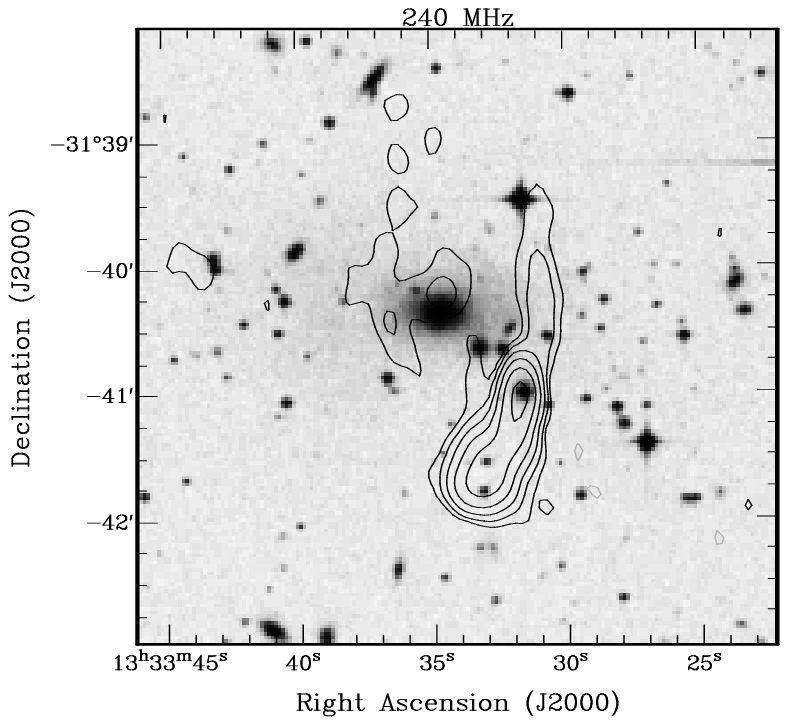}
\includegraphics[angle=0,width=5.9cm]{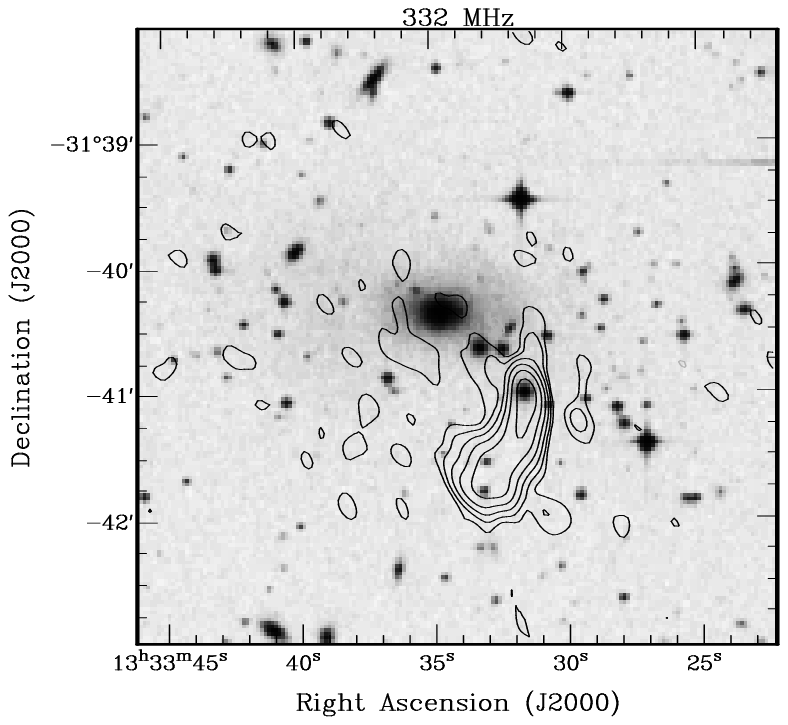}
\includegraphics[angle=0,width=5.9cm]{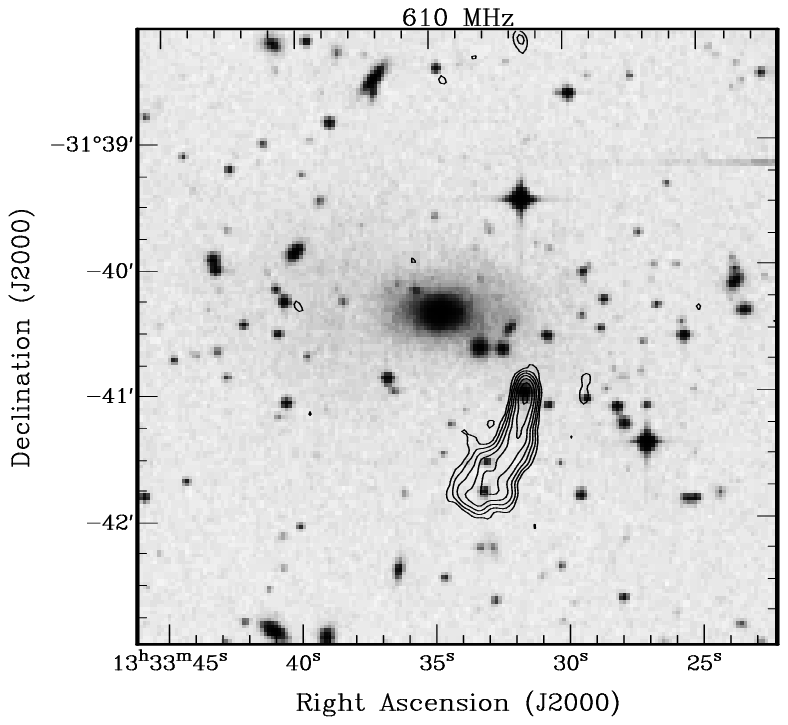}
\caption{Radio contours of the high resolution images of the head--tail radio galaxy 
J1331--3141 at 240 MHz (left), 332 MHz (centre) and 610 MHz (right), overlaid on the
DSS--1 optical frame. The rms
noise is 0.80, 0.30 and 0.15 mJy b$^{-1}$ respectively. Contours levels are -2.4, 2.4, 4.8,
9.6, 19.2, 38.4, 76.8 mJy b$^{-1}$ at 240 MHz, -0.9, 0.9, 1.8, 3.6, 7.2, 14.4,
28.8, 57.6 mJy b$^{-1}$ at 332 MHz, -0.45, 0.45, 0.9, 1.8, 3.6, 7.2, 14.4, 28.8
mJy b$^{-1}$ at 610 MHz. The angular resolution is 
17.4$^{\prime \prime}\times$11.5$^{\prime \prime}$ in p.a. 11$^{\circ}$ 
at 240 MHz, 15.0$^{\prime \prime}\times$8.0$^{\prime \prime}$ in p.a. 26$^{\circ}$ at 
332 MHz and 7.3$^{\prime \prime}\times$5.0$^{\prime \prime}$ in p.a. 7$^{\circ}$ at 610 MHz.} 
\label{fig:headtail}
\end{figure*}
%
%

\section{Radio images}

The GMRT is composed of 30 antennas: 14 are located in 
a $\sim 1.1$ km central compact array and the remaining
are distributed in a roughly Y shaped configuration, with
a maximum baseline of $\sim 25$ km.    
The resulting u--v coverage of the GMRT ensures a good sensitivity
both to compact and to extended sources, and allows deep 
imaging over a wide range of angular resolutions at all 
frequencies. 
Thus we produced images of A3562 at different resolutions,
from few arcsec to tens of arcsec. 
The low resolution images allowed us to enhance the 
low surface brightness emission from the radio halo. Using the 
high resolution images, we could accurately determine the total flux 
density of the tail J1333-3141 and the point--like sources embedded in 
the halo emission. After subtraction of these sources, we could
estimate the halo total flux density. 
The details of all the radio images presented in this paper 
are given in Table \ref{tab:images}.
The values of the flux density of 
the halo refer to the diffuse emission and 
are determined after the subtraction of the head--tail J1333--3141
and of the point--like sources from the total flux density measured
from the images.

%
%
\begin{table*} 
\caption[]{Images details.}
\begin{center}
\begin{tabular}{cccccccc}
\hline\noalign{\smallskip}
Source & RA$_{J2000}$ & DEC$_{J2000}$ & $\nu$  & FWHM, PA & rms   & S$_{tot}$ &LLS $^{1}$  \\
       &  h,m,s& $\circ$,$\prime$,$\prime \prime$ & MHz  & $\prime \prime$, $\circ$ & mJy b$^{-1}$ & mJy & kpc\\
\\
\noalign{\smallskip}
\hline\noalign{\smallskip}
J1333--3141 & 13 33 31.6 & -31 41 02 & 1400 &41.9$\times$35.1, 55 & 0.05 &   109.3 $\pm$ 10.9 & {\bf $\sim$80 $^2$}  \\
            & 13 33 31.6 & -31 41 02 &  610 &7.3$\times$5.0,   7 & 0.15 &   195.7 $\pm$19.6  & $\sim$80  \\
            & 13 33 31.6 & -31 41 02 &  332 &15.0$\times$8.0,  26 & 0.30 &  293.3$\pm$58.7  & $\sim$90 \\
            & 13 33 31.6 & -31 41 02 &  240 &17.4$\times$11.5, 11 & 0.80 &  322.2$\pm$48.3  & $\sim$90  \\
&&&&&&&\\
Radio halo  & 13 33 32.0 & -31 41 00 & 1400 & 41.9$\times$35.1, 55 & 0.05  &20$\pm$2 &  $\sim$475  \\
            & 13 33 32.0 & -31 41 00 &  610 & 30.0$\times$20.0, 7   & 0.15  &90$\pm$9  &  $\sim$470  \\
            & 13 33 32.0 & -31 41 00 &  332 & 41.9$\times$35.1, 55 & 0.25  &195$\pm$39 & $\sim$530  \\
            & 13 33 32.0 & -31 41 00 &  240 & 41.9$\times$35.1, 55 & 0.80  &220$\pm$33 & $\sim$350   \\
\noalign{\smallskip}
\hline
\end{tabular}
\end{center}
$^1$ LLS = Largest Linear Size

{\bf $^2$ Value derived from the 15.0$^{\prime \prime}$ $\times$ 8.0$^{\prime \prime}$ ATCA image 
at 1.4 GHz (V00 and Figure \ref{fig:spixmap_ht} in this paper).}
\label{tab:images}
\end{table*}
%
%

\subsection{High resolution images}\label{sec:low}

In Figure \ref{fig:headtail} we present the high resolution images 
of the centre of A3562, where the head--tail radio galaxy J1333--3141
is clearly visible. Despite the high resolution, residuals of the surrounding 
diffuse emission from the radio halo are visible at 240 MHz and 
332 MHz.

J1333--3141 is associated with the
cluster elliptical galaxy MT 4108 ($b_J=17.25$, v=14438 km s$^{-1}$), 
located at a projected distance of $\sim 1^{\prime}$  
from the cluster dominant cD galaxy (V00). 
A detailed study of this source was carried out 
by V03, using VLA data at 330 MHz, 1.4 , 4.86 and 8.46 GHz,
in addition to older ATCA data at 1.38 GHz and 2.36 GHz
published in V00.

\subsection{Low resolution images}

\begin{figure}
\centering
\includegraphics[angle=-90,width=\hsize]{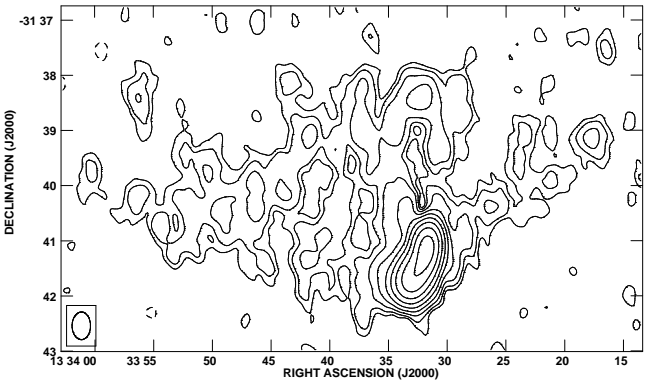}
\caption{Low resolution image at 610 MHz of the central radio halo in A3562.
The restoring beam is $30^{\prime \prime}\times20^{\prime \prime}$.
Contours levels are -0.5, 0.25, 0.5, 1.0, 2.0, 4.0, 8.0, 16.0, 32.0, 64.0 mJy b$^{-1}$.}
\label{fig:halo610}
\end{figure}

At each frequency, the low resolution images of the radio 
emission from A3562 were produced using the full array
with a taper, in order to weight down long baseline data points, 
preserving the total number of visibility points.  
Given the u--v range at each frequency (see Table \ref{tab:obs}), the
best compromise between the low resolution needed and the
sensitivity requested to reliably image and analyse the radio halo
was reached tapering the data to resolutions in the range 
$30^{\prime\prime} - 40^{\prime\prime}$.
For a proper comparison with the VLA 1.4 GHz image of the halo 
in V03, we present here 240 MHz and 332 MHz images made with
the same resolution, i.e. 
$41.9^{\prime \prime} \times 35.1^{\prime \prime}$, in p.a. 55$^{\circ}$.
Since we did not use the 610 MHz data for a point--to--point analysis with the
VLA 1.4 GHz data, in Figure \ref{fig:halo610} we present a 610 MHz image
of the radio halo with the resolution of $30^{\prime \prime} \times 20^{\prime \prime}$,
which better highligths the small scale features of the source.

\begin{figure}
\centering
\includegraphics[angle=-90,width=\hsize]{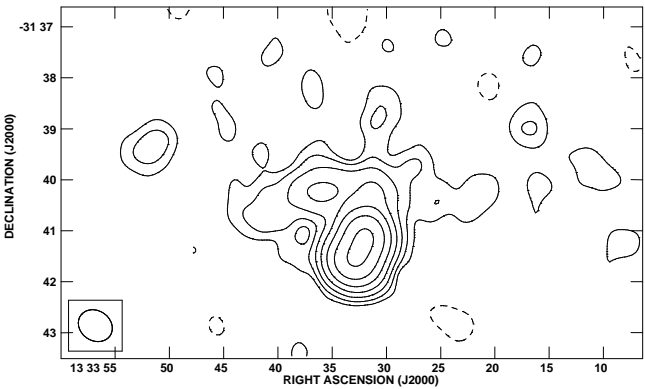}
\caption{Low resolution image at 240 MHz of the central radio halo in A3562.
The restoring beam is $41.9^{\prime \prime}\times35.1^{\prime \prime}$.
Contours levels are -2.4, 2.4, 4.8, 9.6, 19.2, 38.4, 76.8, 153.6, 307.2, 614.4 mJy
b$^{-1}$.}
\label{fig:halo240}
\end{figure}

\begin{figure*}
\centering
\includegraphics[width=\hsize]{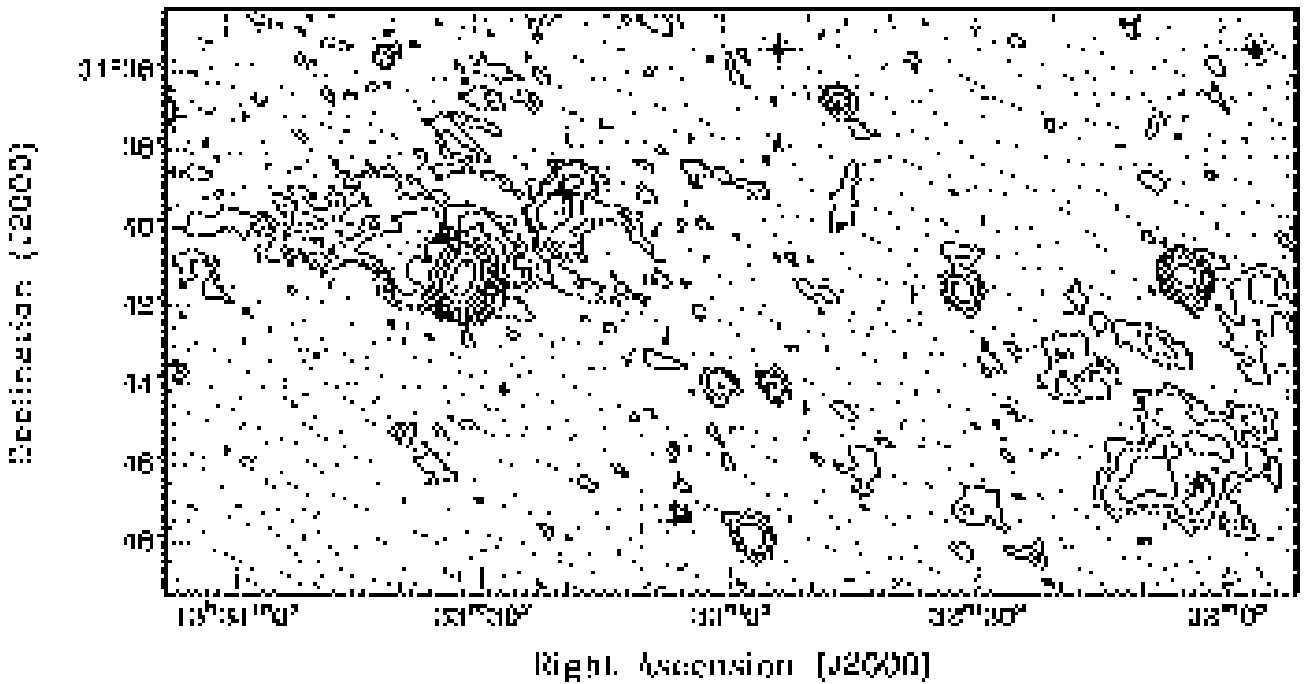}
\caption{Low resolution image at 332 MHz of the region including the
centre of A3562 and the extended radio galaxy J1332--3146a (on the right) overlaid
on the optical DSS--1 frame.
The restoring beam is $41.9^{\prime \prime}\times35.1^{\prime \prime}$.
Contours levels are -1, 1, 2, 4, 8, 16, 32, 64, 128 mJy
b$^{-1}$.}
\label{fig:halo327}
\end{figure*}

Figure \ref{fig:halo240} shows the 240 MHz contours of the radio
halo, and Figure \ref{fig:halo327} shows the radio emission at 332 MHz from the 
whole region of A3562 and SC 1329--313, overlaid on the DSS--1 optical image,
including both the radio halo and the diffuse source J1332--3146a in the SC group.

%
%
%

%
%
%
%

As clear from Figures \ref{fig:halo240}, \ref{fig:halo610} and \ref{fig:halo327}, 
the overall morphology of the radio halo 
at all frequencies shows the same irregular and complex 
structure, which extends in the NE--SW direction.
This is similar to what is found at 1.4 GHz (V03 and Figure \ref{fig:mosaic} 
in this paper).
The largest angular size (LLS) is $\sim 8^{\prime}$ 
($\sim 475$ kpc) at 1.4 GHz and 610 MHz and $\sim 9^{\prime}$ ($\sim
530$ kpc) in the 332 MHz image (see Table \ref{tab:images}).
Note that these values do not include the extension of the wester filament
(see Sect. 3.3).

The noise in the 240 MHz image (Figure \ref{fig:halo240}) is considerably
higher than in the other images (0.80 mJy b$^{-1}$), therefore the lowest contour is
ten times higher than at 610 MHz (Figure \ref{fig:halo610}), and the total
size of the halo at 240 MHz is smaller than at the other frequencies. However, 
there is a clear indication that the region covered by the halo emission 
is consistent with the morphology in Figures \ref{fig:halo610} and \ref{fig:halo327}. 
For this reason, in order to determine the total flux density of the radio 
halo at 240 MHz, we integrated the measurement over 
the whole region of emission covered at the other three frequencies 
(see Table \ref{tab:images}). This value is considerably higher than 
the one derived integrating only the portion of emission 
included in the 3$\sigma$ level (from RA=13$^{h}$33$^{m}$20$^{s}$ 
to 13$^{h}$33$^{m}$45$^{s}$), in support of the presence of positive
residuals in this region. In particular, the flux density within the
$3\sigma$ level is S$_{tot}\sim 170$ mJy, i.e. 
60 mJy less than the value given in Table 2. 
For comparison, the flux density
measured in regions of similar size outside the cluster centre, 
is of the order of the image noise. As a further check of our
estimate of the radio halo size at 240 MHz, we integrated the flux
density over an even wider region, obtaining flux density
values consistent with the one given in Table 2 (within the errors).

The linear size of this radio halo is among the
smallest found thus far. This is not unexpected, since giant radio 
halos (LLS$>$1 Mpc) are usually found in massive and X--ray 
luminous galaxy clusters, while A3562 does not show extreme X--ray
properties (see Section 1). 

%
%
%
%
%

The diffuse source J1332--3146a is visible in the bottom--right
part of Figure \ref{fig:halo327}. This source is associated with
the brightest galaxy in SC 1329--313.
Its extended emission at 332 MHz is strongly elongated to the North--East
direction and points toward the centre of A3562. 
The full resolution images at 240 MHz and 332 MHz of the region of
the J1332--3146a source (not presented here) show no sign of the compact
component associated with the optical galaxy nucleus and detected at
610 MHz (image not shown here) and at higher frequency (V00, G04). 
This suggests that the radio nucleus of the optical counterpart is self--absorbed at 
low frequencies, and that the extended radio emission is not connected to the 
present activity of the AGN. We also note that the 
radio spectrum of J1332--3146a is steep, with a spectral index 
between 240 MHz and 1.4 GHz (G04) $\alpha_{240MHz}^{1.4GHz}=1.2 \pm 0.1$.

\subsection{General comments on the radio emission in A3562}

Our low resolution images show that the radio emission in A3562 is 
characterised by some interesting and peculiar features, and suggest 
some  connection between the centre of A3562 and the region of 
SC 1329--313. The radio halo has an irregular shape, with a filament of emission 
pointing South--West, toward SC 1329--313. This is particularly clear 
at 332 MHz (Fig. 4) and at 1.4 GHz (V03, and also Fig. 9 in this paper). 
The 332 MHz image in Fig. 4 shows also that J1332-3146a is extended 
in the direction of A3562, and that positive residuals (at the 3$\sigma$ level,
and higher) are present between these two regions. These features were noted 
also at 1.4 GHz (G04), and the combination of these two pieces of information
is suggestive of the possibility that very low brightness emission, undetectable 
with the present radio instruments, permeates the region between the centre of
A3562 and SC 1329--313.

\section{Spectral index images and integrated spectra}

Spectral index imaging and integrated radio spectra are powerful tools in
our understanding of the origin of the radio halo emission and of the re--acceleration
processes on cluster scales. However, only very recently
it has become possible to make spectral index maps of radio halos, and to fill 
the low frequency domain ($\nu < 1.4$ GHz) in their integrated spectra. 
In particular, at present spectral index maps are available only for a couple of 
cluster radio halos (Feretti et al. 2004, and ref. therein), 
while a good total spectrum is available 
only for Coma--C, at the centre of the Coma cluster (Thierbach et al. 2003),
and for the central region of the radio halo in the cluster 1E 0657-56
(Liang et al.~2000).

Using the data presented in this paper
and those published in V00, V03 and G04, we produced the images of the spectral 
index distribution for the radio halo  ($\S \ref{sec:spechalo}$) and the head--tail 
radio galaxy J1333--3141 ($\S \ref{sec:spectail}$).
Moreover, we derived the total radio spectrum down to 240 MHz for both sources. 

\subsection{The radio halo}\label{sec:spechalo}

In order to determine the distribution of the spectral index over the radio
halo, we compared the GMRT image at 332 MHz (Fig. \ref{fig:halo327}) and 
the VLA image at 1.4 GHz (Fig. \ref{fig:mosaic}), produced with the 
same cell size and restoring beam (FWHM=$41.9^{\prime \prime} \times 
35.1^{\prime \prime}$, in p.a. 55$^{\circ}$). The images were aligned 
using the AIPS task HGEOM, clipped at a 3$\sigma$ level and 
combined to create the spectral index map within the Synage++ package 
(Murgia \cite{murgia01}).  
We note that the shortest baseline is the same at 332 MHz 
(Table \ref{tab:obs}) and 1.4 GHz (V03).

%
%
%
\begin{figure*}
\centering
\includegraphics[angle=0,width=10cm]{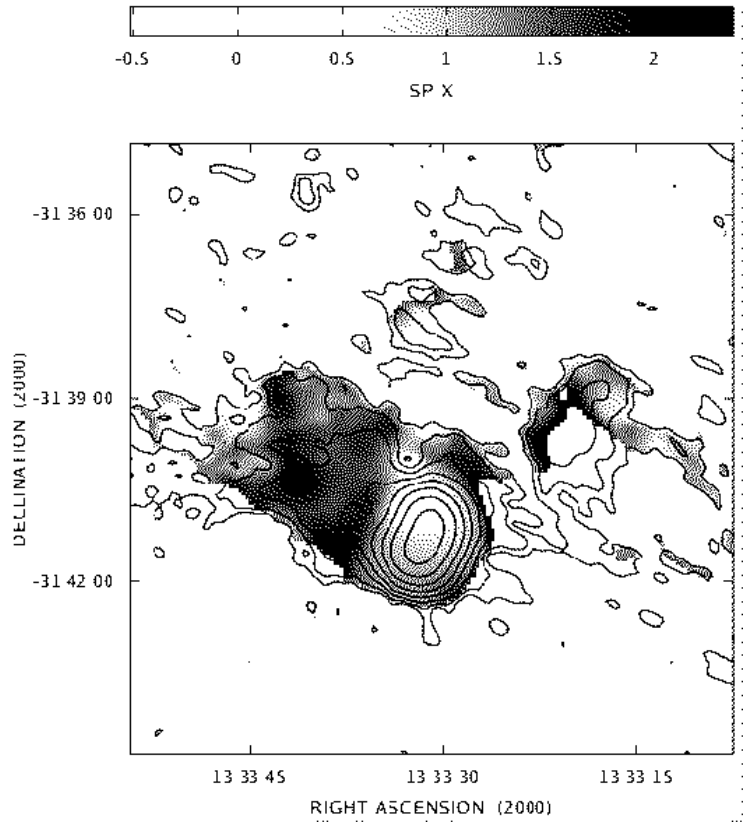}
\caption{Color scale image of the spectral index distribution over the radio halo 
between 332 MHz and 1400 MHz, as
computed from images with a restoring beam of $41.9^{\prime \prime}
\times 35.1^{\prime \prime}$, in P.A. 55$^{\circ}$. Overlaid are the 
GMRT 332 MHz radio contours
at levels -1, 1, 2, 4, 8, 16, 32, 64, 128 mJy b$^{-1}$.
}
\label{fig:spixmap}
\end{figure*}
%
%

Figure \ref{fig:spixmap} shows the spectral index image of the halo
(colours) with 332 MHz contours overlaid. 
It is clear that the region of J1333--3141 has a flat to normal spectrum, with
$\alpha_{332MHz}^{1.4GHz}$ in the range $\sim 0 \div0.7$. The spectrum of the radio
halo is steep, with an average value $\alpha_{332MHz}^{1.4GHz} \simeq 1.5 \pm 0.1$ and
knots steepening up to $\alpha_{332MHz}^{1.4GHz} \simeq 2.0 \pm 0.2$.
Inspection of Figure \ref{fig:spixmap} suggests that there is a clear separation
between the region dominated by the head--tail and the region
dominated by the halo, i.e. 
there is a sharp transition in the spectral index distribution of these two
components, being the latter substantially steeper.  
%
%
\begin{figure}
\centering
\includegraphics[angle=0,width=7.5cm]{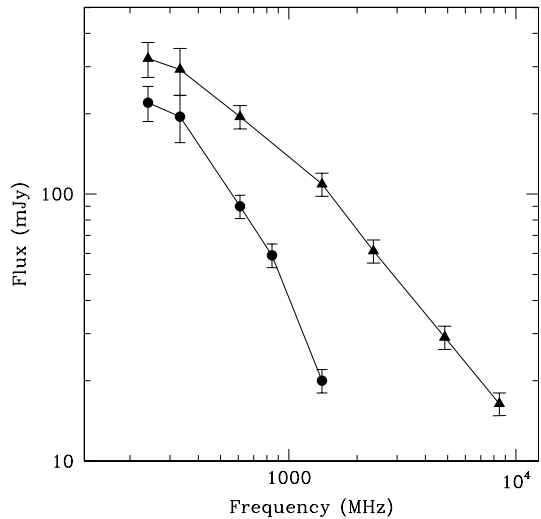}
\caption{Radio spectra of the halo source (points) and of
the head--tail J1333--3141 (triangles) in A3562. 
The 1.4 GHz flux density values are taken from V03. The
843 MHz value for the halo is from V00. For the head--tail, the 
2.36 GHz flux density is taken from V00, and the 330 
MHz, 4.86 GHz and 8.46 GHz values are from V03}
\label{fig:spectrum}
\end{figure}
%
%

We determined the integrated synchrotron spectrum of the radio halo in the frequency 
range 240 MHz--1400 MHz, shown in Figure \ref{fig:spectrum} (points; see also Section
\ref{sec:model} for discussion), using the values 
of the flux density reported in Table \ref{tab:images} and the MOST 843 MHz 
image in V00, who reported a flux density of S$_{843MHz}=59\pm6$ mJy for the halo emission. 
The radio spectrum of the halo appears to be very steep between 1400 MHz and 843 MHz,
with a spectral index $\alpha^{1400MHz}_{843MHz}$ in the range $\sim 1.9 - 2.3$ (V03).
Between 843 MHz and 332 MHz the spectral index is $\alpha^{843MHz}_{332MHz} \sim 1.3 \pm 0.2$, 
and below 332 MHz the spectrum flattens to $\alpha^{332}_{240} \sim 0.4\pm 0.7$.

\subsection{The head--tail radio galaxy J1333--3141}\label{sec:spectail}

%
%
\begin{figure}
\centering
\includegraphics[angle=0,width=7.5cm]{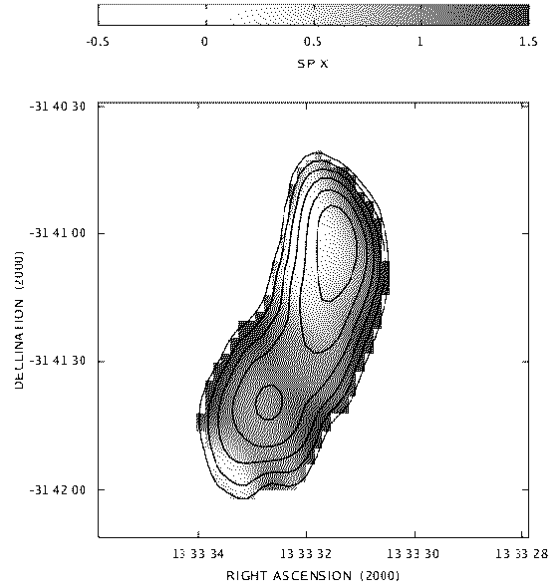}
\caption{Color scale image of the spectral index distribution over the 
head--tail radio galaxy J1333--3141, between 332 MHz and 1380 MHz, as
computed from images with a restoring beam of $15.0^{\prime \prime}
\times 8.0^{\prime \prime}$, in P.A. 26$^{\circ}$.  Overlaid are the ATCA 
1380 MHz radio contours
at levels -0.5, 0.5, 1.0, 2.0, 4.0, 8.0, 16.0 
mJy b$^{-1}$
}
\label{fig:spixmap_ht}
\end{figure}
%
%

In Figure \ref{fig:spixmap_ht} we show the spectral index image
of J1333--3141 (colours), obtained by comparison of the GMRT 
332 MHz image and the ATCA map at 1.38 GHz (contours from V00), produced 
with the same cell size and restoring beam (FWHM=$15.0^{\prime \prime} 
\times 8.0^{\prime \prime}$, in p.a. 26$^{\circ}$). 
As described for the radio halo in $\S \ref{sec:spechalo}$, 
the maps were aligned and clipped at a 3$\sigma$ level before combining them. 

The spectral index distribution clearly shows that we are dealing with a 
tailed radio galaxy. The flattest region ($\alpha_{1.38GHz}^{332 MHz} \sim 0$) 
is coincident with the core region imaged at high resolution and high frequency in V03,
and the significant transverse steepening corresponds to the tail beginning
(the peak in the 1.38 GHz image). 
The spectral index smoothly steepens along the tail up to a value of $\sim 1$.

The total synchrotron spectrum of the source J1333--3141 
in the frequency range 240 MHz--8.46 GHz is shown in Figure \ref{fig:spectrum} (triangles).
The spectrum can be described as a power law with $\alpha_{610 MHz}^{8.46 GHz}=0.96 \pm 0.04$,
with a flattening at frequencies $\nu < 332$ MHz.
Fig. \ref{fig:spectrum} also highlights the well known different spectral 
shape of halo sources and extended radio galaxies at frequencies 
$\nu > 330$ MHz (e.g. the halo source Coma--C, Thierbach et al. \cite{thierbach03},
and low luminosity radio galaxies, Parma et al. \cite{parma02}).

\section{Origin of the radio halo}

The origin of the diffuse radio emission found in galaxy clusters 
is still an open issue. A promising possibility for the explanation of 
radio halos is provided by electron acceleration due to
merger driven turbulence (e.g. Schlickeiser et al. \cite{schlikeiser87}; 
Brunetti et al. \cite{brunetti01a}, \cite{brunetti01b}; Petrosian \cite{petrosian01}).

%
%
%
\begin{figure}
\centering
\includegraphics[angle=0,width=8cm]{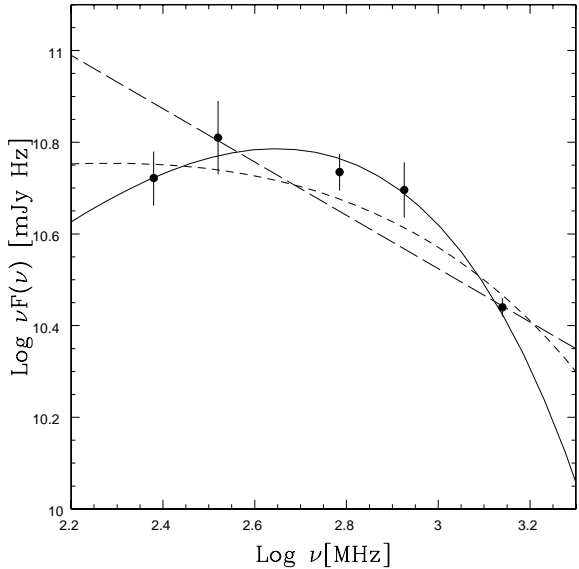}
\caption{
Energy spectrum of the radio halo in A3562 compared with the expectations from three models
for the origin of the emitting electrons.
The {\it long dashed line} represents the emission from a power law energy distribution
of the emitting electrons; this model is normalized at the 332--1400 MHz emission.
The {\it short dashed line} is a turbulent--acceleration model of electrons which
are accelerated
and emit in a region which is larger than the typical scale of variation of the
magnetic field intensity (the slope of the model is similar to what is reported 
in Brunetti et al. \cite{brunetti04}, Fig.~21).
The {\it solid line} is obtained assuming that particles are accelerated 
with a fixed 
acceleration efficiency and emit in a region of constant magnetic field 
intensity. 
}
\label{fig:spectrum_model}
\end{figure}
%
%

During cluster mergers, the infall of the minor 
subcluster through the main one 
generates turbulent velocity fields in the ICM.
This has been observed in the hydrodynamical 
simulations of merging galaxy clusters,  
in the case of both head--on and off--axis collisions
(Roettiger et al. \cite{roettiger97}; 
Ricker \& Sarazin \cite{rs01}).
Turbulence might be powered by the developing
of large scale instabilities at the head of the
colliding subcluster and by the dark matter oscillations
which take place during a cluster merger.
The large scale turbulent eddies diffuse through the cluster volume
and decay into smaller scale turbulence which may accelerate relativistic
particles. A fraction of the energy flux of the fluid turbulence is 
likely channelled into magneto--hydrodynamical (MHD) waves, 
e.g. fast magnetosonic (MS) waves and 
Alfv\'en waves (e.g. Eilek \cite{eilek79}; Fujita, Takizawa \& Sarazin 
\cite{FTS03}; Brunetti et al. \cite{brunetti04}).

Unfortunately the limited dynamic range of present 
numerical simulations does not allow a detailed investigation of
the connection between cluster mergers and
particle acceleration, in particular the cascading of the turbulent 
eddies toward the smaller scales and the generation of MHD modes.
On the other hand a simplified Montecarlo approach to the problem
allows a statistical investigation on the connection between mergers 
and  particle acceleration: a first attempt in this
direction has been recently developed by Cassano \& Brunetti (2005).
Following those calculations, mergers between subclumps with mass 
ratio in the range 2--5 are the responsible for the injection of
the bulk of fluid turbulence in galaxy clusters and are expected
to be those which might activate giant radio halos in the most
massive clusters.
Although these calculations are based on the simplified assumption
that mergers are essentially head--on collisions, the main results
should not be sensitive to the details of the impact parameter, 
provided that the subclump goes across the
innermost regions of the massive cluster (i.e. within 0.2--0.3 
virial radii).

The spectrum of the electrons accelerated by turbulence has a maximum energy
which is due to the balance between radiative losses and acceleration terms.
The main feature of the synchrotron spectrum emitted by these electrons
is the steepening at high radio frequencies, provided that the maximum accelerated 
energy is of the order of that of the radio emitting electrons.

\subsection{Study of the integrated radio spectrum}\label{sec:model}

In Figure \ref{fig:spectrum_model} we report the spectrum of the radio halo 
in A3562 (see Section \ref{sec:spechalo}) compared to the expectations 
from different models for the origin
of the radio emitting electrons (see figure caption). Note that the spectrum is
given as $\nu - \nu F(\nu)$. 

Despite the relatively large error bars at lower frequencies, the high number
of data points and their statistical trend 
strongly constrain the current models for the formation of this radio halo.
It is clear that the observations are not consistent with a simple power law
spectrum (e.g., as expected from secondary models; 
Dennison \cite{dennison80}; Blasi \& Colafrancesco \cite{bc99}).
The A3562 halo is among the few ones which show a steepening of the
radio spectrum with frequency: this provides evidence for diffuse particle 
acceleration in the emitting region.
In the specific 
particle turbulent--acceleration model, which seems to better describe 
the observed spectral behaviour (solid line), we assume that the electrons are 
accelerated and emit in a region where the magnetic field strength is 
approximately constant (reasonably assumed of the order of $\mu$G).
This implies that the scale of the diffuse radio source is comparable to 
(or smaller than) the scale of the variation of the field intensity in the ICM.
The magnetic field in the ICM is believed to be amplified during the cluster 
formation mostly due to dynamo amplification and shear flows and thus it 
should decrease with distance from the cluster centre. Detailed MHD numerical 
simulations show that the radial profile of the field intensity in massive 
clusters is expected to be almost flat in the core region, and it rapidly
decreases with distance (Dolag et al. \cite{dolag02}).
The average size of the halo in A3562 is of the order of the core radius and this
is consistent with the above scenario in which particles are emitting in a region
with relatively uniform field intensity.

%
%
%
\begin{figure*}
\centering
\includegraphics[angle=0,width=\hsize]{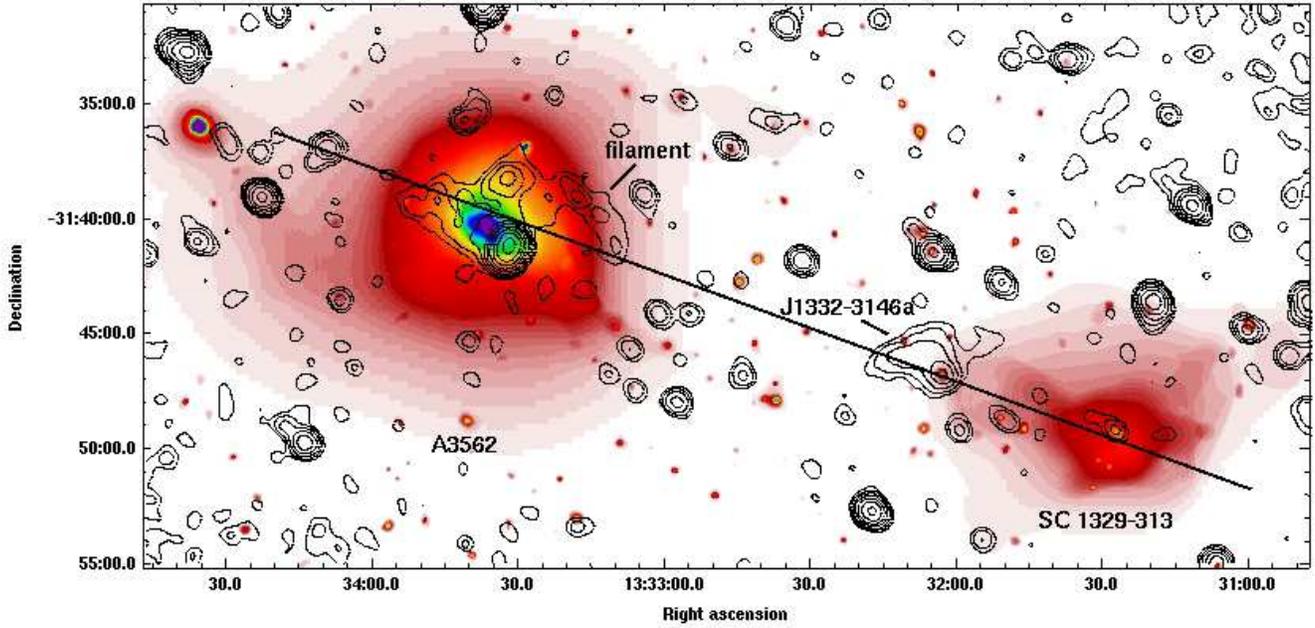}
\caption{VLA--1400 MHz contours of the radio emission from A3562 and SC\,1329--313
superimposed on the XMM--Newton mosaic in the 0.8--2 keV band. 
The radio levels are -0.15, 0.15, 0.3, 0.6, 1.2, 2.4, 4.8, 9.6, 19.2, 38.4, 76.8 and 153.6 
mJy b$^{-1}$. The restoring beam is FWHM=$41.9^{\prime \prime} \times 35.1^{\prime \prime}$, 
in p.a. 55$^{\circ}$.} 
\label{fig:mosaic}
\end{figure*}
%
%

\section{X--radio analysis}\label{sec:X-radio}

In order to complete our analysis on the A3562 radio halo and to
understand its origin and connection with the cluster merger history, we 
performed a composite X--radio study by using  (a) the radio image at 
1.4 GHz first presented in V03; 
(b) XMM--Newton data first presented in F04; (c) {\it Chandra}
observations carried out in ACIS--I with an exposure of $\sim$ 20 ksec 
(Ettori et al. in preparation).

\subsection{X--radio morphological comparison}

Figure \ref{fig:mosaic} shows the overlay of the radio contours 
at 1400 MHz (V03) on the XMM--Newton mosaic of the A3562 region, obtained by F04 
in the 0.8--2 keV band.

\begin{figure}
\centering
\includegraphics[angle=0,width=\hsize]{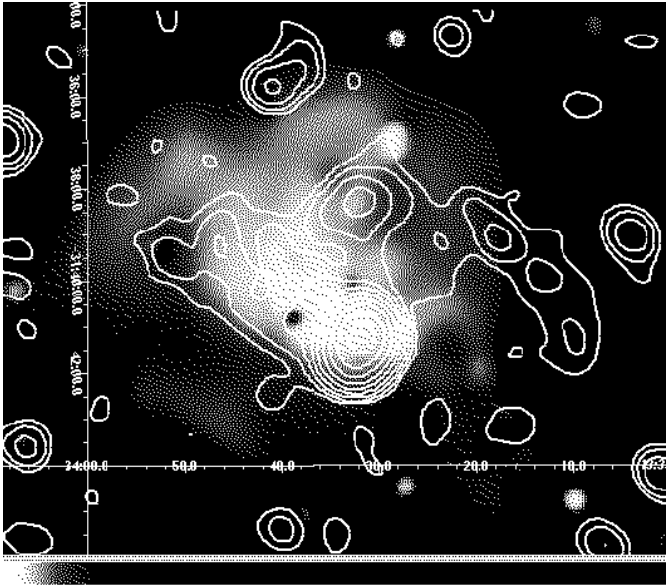}
\caption{VLA-1400 MHz image of the radio halo at the centre of A3562, overlaid 
on the XMM--Newton pseudo--entropy map of the cluster. 
Radio levels are 0.00015 $\times$ (-1,
1, 2, 4, 8, 16, 32, 64, 128, 256, 512, 1024) Jy b$^{-1}$ and the resolution
is $41.9^{\prime \prime} \times 35.1^{\prime \prime}$.}
\label{fig:entropy}
\end{figure}

\begin{figure}
\centering
\includegraphics[angle=0,width=\hsize]{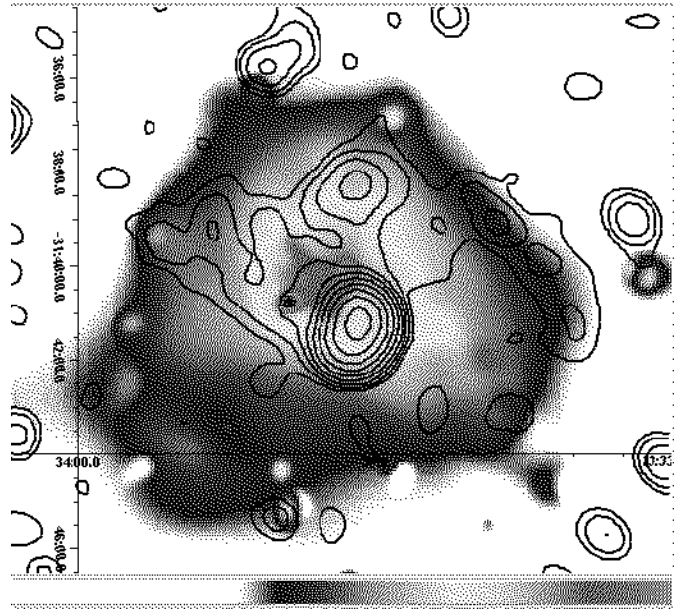}
\caption{VLA 1400 MHz image of the radio halo, overlaid on 
the pressure map of A3562. Radio levels are 0.00015 $\times$ (-1,
1, 2, 4, 8, 16, 32, 64, 128, 256, 512, 1024) Jy b$^{-1}$ and the resolution
is $41.9^{\prime \prime} \times 35.1^{\prime \prime}$.}
\label{fig:pressure}
\end{figure}

In Figures  \ref{fig:entropy} and \ref{fig:pressure} we compare 
the radio emission of the halo at 1400 MHz with the 
pseudo--entropy and pseudo--pressure maps of A3562, derived by F04 
and defined as $S=T / \sqrt[3]{I}$ and and $P=T \times \sqrt{I}$ respectively
(Churazov et al. \cite{churazov03}; Briel et al. \cite{briel03}), where $I$ is the surface
brightness in the 0.8--2 keV band and $T$ is the temperature. 
Figure \ref{fig:mosaic} clearly indicates that A3562 and SC\,1329--313
``feel each other'' both in the radio and X--ray bands. 
In particular:
 
 \noindent i) the X--ray brightness of A3562 is slightly elongated toward SC\,1329--313, 
whose emission strongly points to the centre of A3562; 

\noindent ii) the weak filament of radio emission at the western end of the halo points toward 
SC\,1329--313, while the extended emission of J1332--3146a (located in 
SC\,1329--313) is strongly elongated 
toward A3562;

\noindent iii) the radio halo extends mostly along the North--East direction.

\noindent We point out that the radio and X--ray emission in A3562, the radio galaxy
J1332--3146a and the X--ray emission of SC\,1329--313 are almost perfectly
aligned, as indicated from the line in Figure \ref{fig:mosaic}.

Figures \ref{fig:entropy} and \ref{fig:pressure} zoom into the A3562 centre.
It is noteworthy that the pressure, the entropy and the halo radio emission
extend to North--East with respect to the cluster centre. The inner region
of the pseudo--entropy distribution shows two tails toward North. Hereinafter,
we will refer to these features as to the pseudo--entropy tails.
The South--Eastern part of the radio halo has a sharp edge, and closely follows 
the morphology of the entropy map, shown in Figure \ref{fig:entropy}.   
From Figure \ref{fig:pressure}, we also note that the western radio filament 
pointing South--West remarkably follows (at least in projection) the region of 
enhanced gas pressure commented in F04.

%
%
%

%
%
%
%

\subsection{Point--to--point comparison}\label{sec:chandra}

\begin{figure}
\centering
\includegraphics[angle=0,width=\hsize]{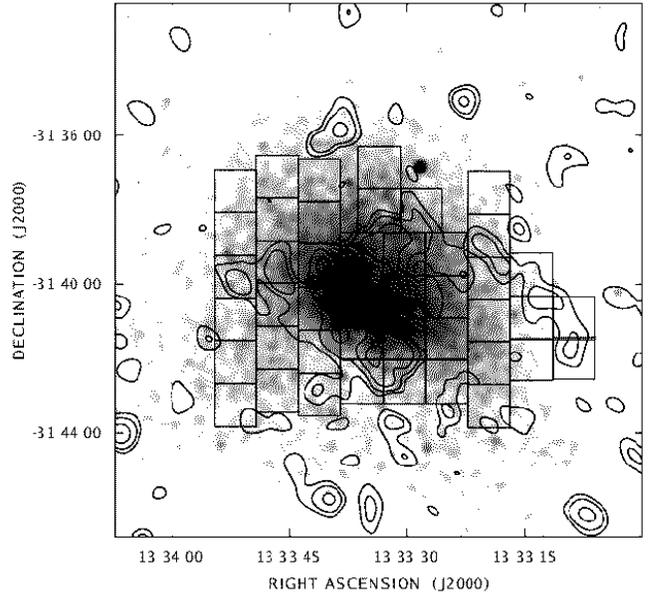}
\caption{Grid used for the comparison of the radio and X--ray images of the cluster A3562. 
The size of each grid cell is 70$^{\prime \prime}$ $\times$ 70$^{\prime \prime}$. The grid is 
overlaid on the {\it Chandra} X--ray surface brightness image. 
Contours represent the VLA-1400 MHz image of the halo, after subtraction of the central 
head--tail radio galaxy J1333--3141. Radio levels are 
0.00015 $\times$ (-1, 1, 2, 4, 8, 16, 32, 64, 128, 256, 512, 1024) Jy b$^{-1}$. 
The angular resolution of the radio map is $41.9^{\prime \prime} \times 35.1^{\prime \prime}$.
The X--ray image has been obtained from 
the {\it Chandra} raw image (resolution 2$^{\prime \prime} \times 2^{\prime \prime}$) 
after a smoothing with a box of 60$^{\prime \prime}$ width.}
\label{fig:grid}
\end{figure}

We performed a point--to--point analysis of the radio (flux density at 1.4
GHz and 332--1400 MHz spectral index) and the X--ray (surface brightness,
temperature, entropy, pressure) quantities, using 
both {\it Chandra} and XMM--Newton data.
Following Govoni et al. (\cite{govoni01b}) we constructed 
a grid covering the cluster region and determined the mean radio 
and X--ray quantities for every grid cell, 
as well as the root--mean--square (rms), which can be assumed 
as an estimate of the error. 
The whole analysis has been carried out using the IDL and the Synage++ packages.
In order to perform a proper comparison, we used an image of the halo 
after subtraction of the central head--tail radio galaxy J1333--3141. 
In Figure \ref{fig:grid} we show the radio halo 1.4 GHz emission (after subtraction 
of the head--tail source) as contours, superposed 
on the {\it Chandra} X--ray surface brightness image.
The Figure also reports the grid used for the analysis, which covers the whole
region of X--ray emission. The size of each grid cell is 
$70^{\prime \prime} \times 70^{\prime \prime}$, in order to select
statistically independent regions.

As noted in the comparison between the radio halo and
the XMM--Newton images (Sec. 6.1), we find a strong spatial 
correlation between the radio halo emission and the inner region 
of the emission from the X--ray plasma.

We investigated the presence of
correlation between radio and X--ray quantities performing a Spearman test, which
evaluates a rank correlation coefficient, $r_s$, on given data arrays.
The absolute value of $r_s$ can range between 0 (no correlation is present)
and 1 (perfect correlation).
The test returns also $P$, the two--sided significance of the agreement
of the rank correlation coefficient, $r_s$, with the null--hypothesis
of no-correlation between the data arrays. A small value of $P$ indicates
a significant (anti--)correlation if $r_s$ is (negative) positive.

\begin{figure}
\centering
\includegraphics[angle=0,width=\hsize]{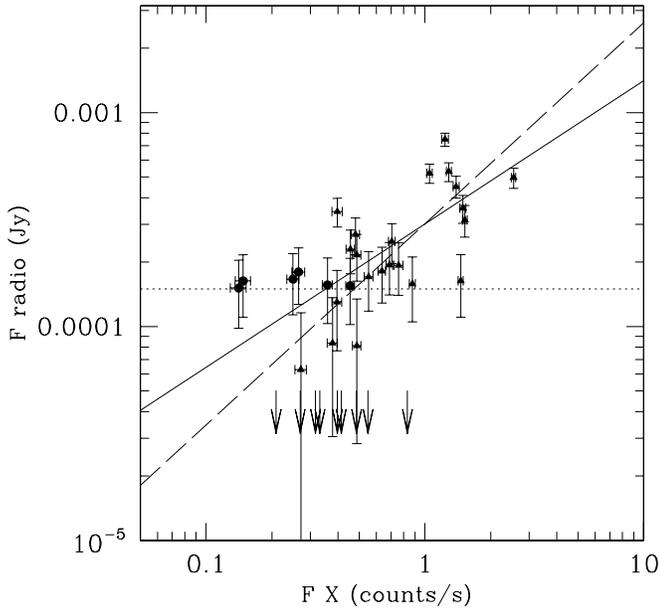}
\caption{Relation between the radio flux density at 1.4 GHz and the {\it Chandra} 
X--ray flux for the cluster A3562. Data points represent the mean flux density in each 
cell of the grid constructed over the cluster region (Fig.\ref{fig:grid}). 
The error--bars are the rms of flux distribution. The filled dots indicate the values 
derived for the grid cells containing the weak filamentary structure of the radio halo 
extending toward South--West. The upper limits represent cells with no detected
radio emission above the 1$\sigma$ level. The dotted line is the radio 
3$\sigma$ level. The solid line is the best fit of the data including all the boxes. 
The long--dashed line is the best fit obtained excluding the western filament.}
\label{fig:correlation}
\end{figure}

The most significant correlation is given by the {\it Chandra} X--ray surface 
brightness versus the radio flux density at 1400 MHz (see Figure \ref{fig:correlation}).
In the Figure filled triangles 
represent the data from the cells with radio emission; filled dots represent
the data from the cells which cover the western filament. For completeness
we show also some upper limits (not included in the statistical analysis), 
which represent cells with no detected radio emission above the 1$\sigma$ level. 
For these two quantities   
we found $r_s=+0.65$; the probability
that the relative distribution of values is not correlated is 
P $= 1.5 \times 10^{-4}$.
In this case the weighted best fit is given by logF$_{radio} = (0.67 \pm 0.12)$logF
$_X - (3.52 \pm 0.05)$, which is consistent with a sub--linear correlation,
as found in some other cases (Govoni et al. 2001b).
A correlation between radio flux and X--ray brightness is also confirmed
by the analysis of the XMM--Newton data.
It is worth noticing that if we exclude the western filament of the halo
from the statistical analysis, the significance of the correlation slightly
decreases to $r_s=+0.60$ (P$=3.3 \times 10^{-3}$), meaning that the
filament plays a marginal role in the overall significance of the 
F$_{radio}$--F$_{X}$ correlation, while the slope of the correlation steepens
from sub--linear to linear (logF$_{radio} = (0.94 \pm 0.13)$logF
$_X - (3.52 \pm 0.05)$), as clear from Fig. \ref{fig:correlation}.

Moreover, higher radio fluxes are measured in regions
with lower values of X--ray gas temperatures and entropy levels 
($r_s \sim -0.6$) and higher gas pressure estimates ($r_s \sim +0.6$). 
We detect weak evidence ($\mid r_s \mid \le 0.4$) for a spatial correlation between 
the spectral index image of the radio halo (Fig. 5) and the projected X-ray
quantities.
We note that these results do not depend on the choice of the position
of the boxes to cover the cluster emission (see  Fig.12),
providing different configurations very similar results.

%
%
%
%
%

%
%
%
%

\section{Cluster merger and origin of the diffuse radio emission}

Given the high quality of the X--ray and radio data and the 
evidence for diffuse particle acceleration from the integrated 
radio spectrum (Sect.5.1), 
in this Section we attempt to explain the origin of the diffuse
radio emission as due to electron acceleration driven by
the turbulence generated in the ICM during the merger between 
A3562 and SC\,1329--313.

The efficiency of the turbulence cascading and of the 
vwave--particle coupling depends on many physical quantities,
which are basically unknown (e.g. the magnetic field strength and
topology, the spectrum of the MHD turbulence).
However, it has been recently pointed out that, independent
of the details of the turbulence injection process,
MS waves with scales $<$100 kpc may 
efficiently accelerate fast electrons in the ICM
up to the energies ($\gamma \sim 10^4$) required to emit the
synchrotron radiation observed in the radio band
(Cassano \& Brunetti \cite{cb05}).
The decay time scale of the MS waves is approximately 
$\tau_{kk} \sim  v_{_{MS}}/k v_k^2$ (e.g., Yan \& Lazarian \cite{yl04}),
where $v_{MS}$ is the magnetosonic velocity, 
$v_k$  \footnote{We used $v_k^2 \sim {\eta_{_{MS}}{E_{th}\over 2\rho}}$,
$\rho$ = density of the thermal ICM, using a Kraichnan spectrum.}
is the velocity
of the turbulent eddies and $k$ is the wavenumber.
Given the physical conditions in the ICM of A3562, it is:
 
\begin{equation}
\tau_{kk} ({\rm Gyr}) \simeq 0.7 {{ L}\over{ {\rm Mpc} }} 
\eta_{_{MS}}^{-1}
\label{taukk}
\end{equation}

\noindent
where $\eta_{_{MS}} = E_{\rm MS}/E_{\rm th}$ is the ratio between 
the energy in the form 
of MS waves and the thermal energy in the emitting region 
($\eta_t \sim 0.2 - 0.9$ in Cassano \& Brunetti \cite{cb05} is a few times 
$\eta_{_{MS}}$ here), 
and $L$ is the maximum injection scale of the MHD turbulence.
This is comparable to the crossing time of massive subclumps through
the volume of the main cluster and thus it may allow an effective spatial
diffusion of the turbulence over Mpc scales with a fairly uniform 
intensity before being dissipated.

The proposed scenario for the origin of the radio halo in A3562 is
based on the following points.

(1) {\it Merger Kinematics} -- Making use of several arguments,
F04 derived the kinematics of the collision between SC\,1329--313 and A3562.
They suggested that about 1 Gyr ago SC\,1329--313 passed $\sim 500$ kpc
North of the core of A3562 (corresponding to  $\sim$ 0.25 virial radii) 
coming from the East, with a relative velocity of the 
order of 1700 km s$^{-1}$, and it was gravitationally
deflected toward the South--Western direction. Note that the mass
ratio between A3562 and SC\,1329--313 is 3:1 (Ettori et al. 1997).
As a consequence of this passage, the core of
A3562 acquired a velocity component in the North--South
direction and it started sloshing, with an oscillation amplitude
of the order of 200 kpc and a period of $\sim 1$ Gyr.

(2) {\it Magnetic field and Merger} --
The change of the core velocity at the oscillation apogee causes
the escape of low--entropy gas. This  
produces the entropy tails, as discussed in F04 (see also Figure \ref{fig:entropy}
in Section 6.1). 
Since the magnetic field is frozen into the thermal plasma,
we expect that the escape of the gas should flatten the profile of the field strength  
toward the region North of the A3562 core.

(3) {\it Turbulence and particle acceleration} -- 
Another expected by--product of the interaction between SC\,1329--313 and A3562 
is the injection of turbulence in the ICM.
Although the injection mechanism of the turbulence is very complex
and depends on many unknown parameters, we expect that large
scale turbulence may be efficiently injected in the volume swept by 
the head of SC\,1329--313 (i.e. by the self bounded subclump which
is gradually reduced during the crossing of the main cluster
by the effect of the ram pressure stripping)
and in the central region of A3562 by the North--South oscillation of the
core. As shown by analytical arguments (Cassano \& Brunetti 2005) and
by numerical simulations (Ricker \& Sarazin 2001), once injected these
large scale eddies diffuse trough a larger volume so that 
one may reasonably assume that turbulence fills at some
level the region of the core of A3562 and a large fraction of the 
volume between the cores of the two subclusters (actually from the 
bridge toward the north).

The effect of the turbulence on the process of particle acceleration 
and amplification of the magnetic field depends on the strength of 
the turbulence in this region.
Provided that a fraction 
$\sim 0.005-0.1 \times (10^{-3}/n_{\rm th}) (10^8/T)$ 
of the thermal energy of the ICM is channelled
into MHD waves (depending on the specific MHD mode, see Brunetti et al.
2004, and Cassano \& Brunetti 2005),
this may switch on the diffuse radio emission observed near the centre of 
A3562.

\subsection{Self-consistency of re--acceleration}

If electrons are re--accelerated, as indeed suggested by the curved
radio spectrum of A3562 (Figure \ref{fig:spectrum_model}),
similar to Coma--C (Thierbach et al. \cite{thierbach03}),
then the source age can be estimated by the cascading time--scale 
of the turbulence injected on the scales of the instabilities generated
during the merging process. 
An analysis of the entropy map limits the scale of the Rayleigh--Taylor
instabilities, driven by the core oscillation, to 
$\sim 500$ kpc; the corresponding cascading time--scale
(Eq.~\ref{taukk}) is $\tau_{\rm kk} = 0.7 
\left( {{ \eta_{_{MS}}\over{0.5}}} \right)^{-1}$Gyr; this should be
considered a lower limit, since the oscillation of the core (and 
hence the growth of the instabilities) is still on--going.

At the same time, large scale turbulence may be injected at the 
passage of SC\,1329--313 North of the A3562 core.
In this case, the initial scale of the instabilities should be 
of the order of the diameter of the core of SC\,1329--313 and thus 
the cascading time should be $\sim 1$ Gyr.
We note that both these cascading time--scales are of the
order of the time--scale of the merging,
and  of the core oscillation in  A3562 (F04).

Thus the proposed scenario is self--consistent:
there is enough time for the development
of a turbulence cascade, necessary for particle acceleration; 
at the same time, the epoch of turbulence injection is recent
enough to expect that turbulence is not yet completely dissipated
at present.

\subsection{Observational evidence}

The observed correlations between radio and X--ray quantities
(Sects.~\ref{sec:X-radio}) support the proposed scenario.

If turbulence is injected by the passage of SC\,1329--313 through
the cluster volume and by the oscillation of the A3562 core, 
then it should be fairly uniform on several hundreds kpc scale.
Thus, if the energy density of the turbulence is smaller than that of the
thermal plasma, its intensity should not significantly depend on the
X--ray quantities (e.g. temperature and thermal density).
In this case, once the relevant wave--damping processes are taken into account
(e.g., Cassano \& Brunetti \cite{cb05}), 
the efficiency of the particle acceleration due to
MS waves should scale with the temperature as $1/\sqrt{T}$ and thus an 
anticorrelation between the 
non--thermal emission and the higher temperature patches is expected.

On the other hand, since the magnetic field, $B$, is frozen into the ICM, 
higher synchrotron emissivity ($\propto B^2$) is expected in regions with
higher density, $n_{\rm th}$, provided that fast electrons are accelerated in these
regions. The significant correlation between radio and X--ray ($\propto
n_{\rm th}^2$) flux could thus be driven by this effect.
The freezing of the magnetic field into the ICM could also account for the
observed anti--correlation between the radio flux and the entropy, provided
that higher values of $B$ are found in regions with lower entropy.
Similarly, the observed correlation between the radio flux and the thermal
pressure is expected if the measured values of the pressure are mainly 
determined by the values of the gas density.

\subsection{The extent of the radio emission}

The observed extent of the radio emission is slightly
smaller than the entropy tails of A3562. 
Cassano \& Brunetti (\cite{cb05}) calculated that 
cluster mergers may efficiently power 
giant (i.e., $\geq$ Mpc sized) radio halos 
only in the case of massive clusters 
(with M$_{200} \geq 10^{15}$M$_{\odot}$).
This threshold is essentially due to the fact that
the energy density (within 1 Mpc$^3$) of the injected turbulence 
in clusters with masses below this value 
is too small to drive efficient particle acceleration on 
large (Mpc--extended) regions, and thus the cut--off in the
synchrotron spectrum falls below the typical radio--observing frequencies.
A3562 is slightly below this mass threshold
($M_{200} \simeq 5.6\times 10^{14}$M$_{\odot}$),
and the presence of a giant radio halo is
indeed not expected by these calculations.
In this case, however, it might be thought that
diffuse synchrotron emission may be revealed on smaller scales 
in the cluster--core region and toward the North 
of the core, where the strength of the magnetic field is likely
to be stronger (and amplified by the interaction between the 
two subclusters).

The clear steepening of the radio spectrum
(Figure \ref{fig:spectrum_model})
strenghtens this interpretation, since it
proves that the energy of the electrons accelerated in the core region
is just sufficient to emit synchrotron
radiation up to $\leq$GHz frequency.

Given the shape of the observed radio spectrum, we point out that
even assuming that the acceleration efficiency is maintained outside
the radio emitting region, 
a drop in the magnetic field strength by a factor of $\sim 2$
outside this region would cause a sharp decrement of the synchrotron
emissivity at 610 MHz, i.e. a factor of $\sim 7-8$.

\subsection{A radio bridge between the radio halo and J1332--3146a?}

Figure 4 and the 1.4 GHz images presented in Fig. 9 and in G04 show 
outstanding features of extended radio emission. In particular, the
western filament of the radio halo, the extention of J1332--3146a 
in the direction of the A3562 centre, and the presence of positive
residuals of radio emission at 1.4 GHz and 332 MHz are all
perfectly aligned.
Furthermore, the radio/XMM--Newton overlay of Fig. 9 shows that 
the X--ray emission is also aligned in the same direction.
\\
The scenario proposed in this section could account for these
observed signatures, at least at a qualitative level. 
It seems reasonable to expect that the motion of SC\,1329--313 
toward South--West and the gravitational interaction with A3562
have actually produced a ``channel'' of turbulence, which may have 
affected the ICM and relativistic particles over the whole
distance between the group and the cluster.
If this is the case, the energy rate injected into MHD 
modes along the bridge should be roughly of the order of
\gtsim $~~5 \times 10^{-29}$erg s$^{-1}$cm$^{-3}$
(here, for simplicity, we assume MS waves and a reference
value of $B \sim 0.5 \mu$G and $T \sim 5 \times 10^7$K),
which immediately would imply that the fluid 
turbulence in this low density region should be strong, from 
slightly subsonic to mildly supersonic depending on
the efficiency of the radiation process of MS waves.
\\
The radio images suggest that the western filament and the
extension of J1332--3146a are the ``peaks'' of a bridge of
very low brightness extended radio emission, which falls
below the sensitivity limits of the current radio interferometers.
This was noticed also in G04, based on  1.4 GHz data. The new
GMRT data presented here, in particular the 332 MHz image,
further strengthen this possibility.
We note that the spectrum of J1332--3146a is steep, with
$\alpha^{1.4GHz}_{240MHz} = 1.2$, which is consistent with what
is found in cluster--type radio sources such as halos and
relics.

\section{Summary and Conclusions}\label{sec:disc}

In this paper we presented and discussed new low--frequency GMRT 
images
at 240 MHz, 332 MHz and 610 MHz of the recently discovered 
radio halo at the centre of the A3562 cluster.
 
In the following we summarize the most important results of 
these observations, combined with our
previous 1.4 GHz VLA data (published in V03), and
the {\it Chandra} and XMM--Newton X-ray observations carried out by 
Ettori et al. (in prep.) and F04.

\begin{itemize}

\item[1)] A3562 is characterized by extended radio emission
on the cluster scale. Beyond the presence of a radio halo 
at its centre, the radio galaxy J1332--3146a (located South--West
of A3562) shows low brightness extended emission in the direction of 
the radio halo, suggestive of a possible connection between the
two features. This possibility is reinforced by the presence
of positive residuals between these two features, both at 332 MHz
and at 1.4 GHz (G04).

\item[2)] The radio halo in A3562 is among the smallest found
thus far. Its LLS is of the order of $\sim$500 kpc
going from 1.4 GHz to 332 MHz.

\item[3)] The point--to--point spectral index image of the halo
shows a complex structure. Beyond the existence of
a region with an average value of 
$\alpha_{332MHz}^{1.4GHz} \sim 1.5 \pm 0.1$, there are a number of
 knots steepening up to $\sim 2.0\pm0.2$.  Our images also show
 that the spectrum of the 
 electron populations of the radio halo and of the
 tailed source J1333--3141 embedded in it are clearly different.

\item[4)] With all the data available we could derive the
integrated radio spectrum of the halo in the range 240 MHz--1.4 GHz, 
with 5 data points. The spectrum is very steep between 
1400 MHz and 843 MHz, with a spectral index $\alpha^{1400MHz}_{843MHz}$ 
in the range $\sim 1.9 - 2.3$ (V03).
Between 843 MHz and 332 MHz the spectral index is 
$\alpha^{843MHz}_{332MHz} \sim 1.3 \pm 0.2 $, 
and below 332 MHz the spectrum flattens to $\alpha^{332}_{240} \sim 0.4 \pm 0.7$.
We also derived the spectrum for the extended source 
J1332--3146a, and found that it is also steep, with
$\alpha^{1.4GHz}_{240MHz} = 1.2 \pm 0.1 $.

\item[5)] The radio and X--ray images show that A3562 and the group
SC\,1329--313 "feel each other", i.e. the extended features of the two regions
point toward each other in both bands.

\item[6)] A spatial correlation is found between the radio halo and
the quantities derived from the XMM--Newton and {\it Chandra} observations, like
the X--ray surface brightness, the temperature, the pseudo--entropy and 
pseudo--pressure maps. All the observables show an
extension toward North, and a sharp edge South of the core region in A3562.
A significant point--to--point linear positive correlation holds 
between the 1400 MHz flux density of the halo and the X--ray flux 
density. Such correlation is expected, given the correlations observed between the 
halo emission and the pseudo--pressure and --entropy maps of A3562.
Furthermore, the western filament of the radio halo, pointing to South--West,
follows remarkably (at least in projection) a region of enhanced gas
pressure.
\end{itemize}

Our results give further observational support to a scenario of cluster 
merger between A3562 and SC\,1329--313. 
In particular, we discussed the hypothesis 
that the radio halo in A3562 is the result of particle re--acceleration
induced by turbulence injected in the ICM by the passage of SC 1329--313
North of the core of A3562, and by the oscillation of the cluster core in
the North--South direction. The proposed scenario may explain the properties 
of the radio halo, including its relatively small size, the morphology and 
extent  of the pseudo--entropy tails as derived from the XMM--Newton
observations, as well as the statistical correlations between the
radio and X--ray observables. 
\\
We further speculate that the alignment of the western filament of the radio 
halo with the extended emission around J1332--3146a and the positive 
residuals of radio emission between the centre of A3562 and the group 
SC\,1329--313, are also connected to the turbulence induced by this
merger event.

It has been shown that cluster mergers involving massive 
(M$>10^{15}$ M$_{\odot}$) and luminous (L$_X > 10^{45}$ erg s$^{-1}$) 
clusters may develop the energy and turbulence required to produce 
Mpc--scale radio emission, in the form of giant radio halos 
(Cassano \& Brunetti, 2005). On the other hand, it might be expected that 
slightly less massive systems might also produce diffuse radio emission,
though less extended and less powerful than that of giant radio halos.
Indeed, the case of A3562 shows that merger events involving clusters 
with mass M$\sim 6 \times 10^{14}$ M$_{\odot}$ might produce
signatures in the radio band, at the detection limit of the 
intrumentation currently available. Therefore we expect that future 
observational facilities, such as LOFAR and SKA, will substantially
contribute to our knowledge and understanding of the cluster 
merger phenomenon.

\bigskip\centerline\
We thank the staff of the GMRT who have made these
observations possible. GMRT is run by the National Centre for
Radio Astrophysics of the Tata Institute of Fundamental Research.
The paper is also based on observations obtained with XMM--Newton,
an ESA science mission with instruments and contributions
directly funded by ESA Member States and the USA (NASA).
We thank R.Cassano for useful discussions, and the anonymous referee
for very fruitful comments.
G.B. acknowledges partial support from INAF from grant D4/03/15
and from MIUR from grant PRIN2004.
A.F. acknowledges support from BMBF/DLR under grant 50 OR 0207 and
MPG. The present research was carried out with partial support
of the contract ASI--I--R--063--02.

\end{document}